\documentclass[a4paper,11pt]{article}
\usepackage{subfig}
\usepackage{jinstpub} 
\usepackage{lineno}

\title{\boldmath Efficiency measurements of GEM GE1/1 chambers in the upgraded CMS Endcap Muon System using 2023 collision data at $\sqrt{s}=13.6$ TeV}

\author[o]{M.~Abbas,} 
\author[ai]{S.~Abbott,}
\author[t]{M.~Abbrescia,}
\author[j,l]{H.~Abdalla,}
\author[j,m]{A.~Abdelalim,}
\author[j,1]{S.~AbuZeid, \note{Now at Universit\`{a} di Pavia and INFN Sezione di Pavia, Pavia, Italy}} 
\author[ah]{D.~Aebi,}
\author[ad]{A.~Ahmad,}
\author[ad]{W.~Ahmed,}
\author[x]{C.~Aim\`{e},}
% ...

% ...
\affiliation[j]{Academy of Scientific Research and Technology - ENHEP, Cairo, Egypt} %
% ...
\affiliation[l]{Cairo University, Cairo, Egypt}
\affiliation[m]{Helwan University, also at Zewail City of Science and Technology, Cairo, Egypt}
% ...
\affiliation[o]{Karlsruhe Institute of Technology, Karlsruhe, Germany}
% ...
\affiliation[t]{Politecnico di Bari, Universit\`{a} di Bari and INFN Sezione di Bari, Bari, Italy}%
% ...
\affiliation[x]{Universit\`{a} di Pavia and INFN Sezione di Pavia, Pavia, Italy} %
% ...
\affiliation[ad]{National Center for Physics, Islamabad, Pakistan}
% ...
\affiliation[ah]{Texas A$\&$M University, College Station, USA}
\affiliation[ai]{University of California, Davis, USA} %
% ...

\emailAdd{shalini.thakur@cern.ch}

\abstract{The CMS experiment at the Large Hadron Collider employs Gas Electron Multiplier (GEM) detectors, a technology based on gaseous ionization, as one of the muon detectors. The muon spectrometer is being upgraded to handle the increased muon flux in the forward region. This study analyzes muon detection efficiency in the GE1/1 triple-GEM detector, using 2023 proton-proton collision data at $\sqrt{s}=13.6$ TeV. A dataset enriched with muons from Z boson decay, with a total recorded luminosity of 17.8 fb$^{-1}$ has been used for this study. The detection efficiency of 137 GEM detectors are measured using muon trajectories established using other detectors in the tracking and muon systems, without use of the GEM detectors. The average efficiency of 137 GEM detectors is $\sim$93.3$\%$.  A subset of 108 detectors that had no shorts were operated at the nominal HV working point with average efficiency of $\sim$96$\%$. Efficiency is found to be unaffected by the number of p-p interactions per bunch crossing (pile-up).}


\begin{document}
\maketitle
\flushbottom
\clearpage 

\section{Introduction}
\label{sec:intro}
\subsection{The Large Hadron Collider and the Compact Muon Solenoid Detector}

The Large Hadron Collider (LHC)~\cite{a,b} at CERN is the world’s highest-energy particle accelerator, colliding protons at a center-of-mass energy of 13.6 TeV during Run 3 (2022–2026). With instantaneous luminosities reaching 2.5 $\times 10^{34}$ cm$^{-2}$s$^{-1}$, the LHC continues to provide unprecedented datasets for precision Standard Model measurements and searches for new physics. The upcoming High Luminosity LHC (HL-LHC) run~\cite{c,d}, expected to start in 2030, will increase the integrated luminosity by nearly an order of magnitude, demanding significant upgrades to the experiments to maintain high efficiency, radiation tolerance, and triggering capability under challenging pile-up conditions.
\\
\\
The Compact Muon Solenoid (CMS) experiment~\cite{e,f,g} is a general-purpose detector designed to explore the physics program at the LHC with high precision. Its superconducting solenoid provides a 3.8~T magnetic field within which a silicon tracker, an electromagnetic calorimeter, and a hadronic calorimeter measure and identify charged particles, photons, and jets. Outside the solenoid, the muon system, as configured before Run 3 in 2022, consisted of Drift Tubes (DTs), Cathode Strip Chambers (CSCs), and Resistive Plate Chambers (RPCs) embedded in the return yoke, providing efficient muon triggering, reconstruction, identification, and precise transverse-momentum measurements up to pseudorapidity $|\eta|<2.4$. To sustain performance in the forward region under HL-LHC conditions, a new generation of Gas Electron Multiplier (GEM) detectors has been introduced.
\\
\\
The first system of GEM detectors, GE1/1~\cite{h,i}, was installed during the LHC Long Shutdown 2 (LS2) (2018–2022)~\cite{j} and they cover the pseudorapidity range $1.55<|\eta|<2.18$. The GE1/1 chambers complement the ME1/1 CSCs by improving muon reconstruction, enhancing the momentum resolution of muons, and reducing Level-1 trigger rates. Figure 1 shows the layout of the upgraded CMS muon system, including GE1/1.
\\
\begin{figure*}[htb!]
\centering
\includegraphics[width=0.65\linewidth]{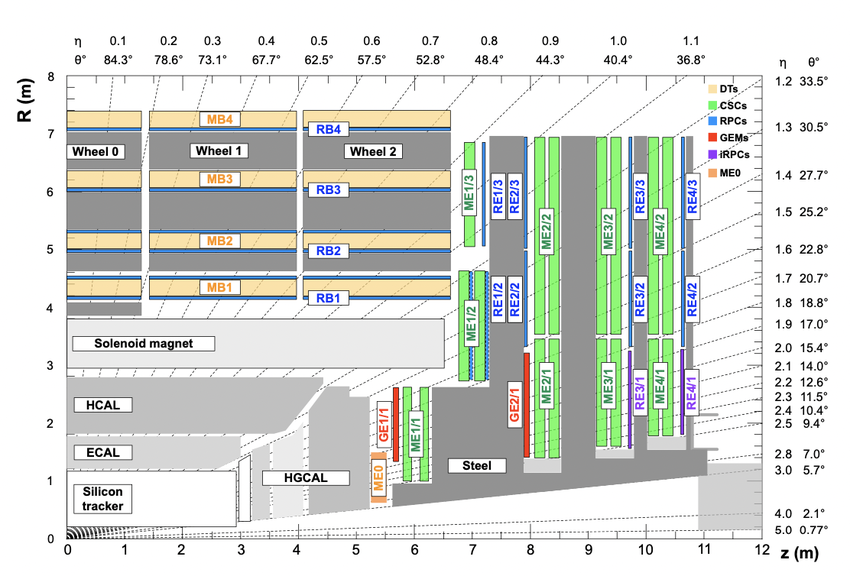}
\caption{A longitudinal view of the CMS muon system for the HL-LHC upgrade. The GEM detectors are shown in red and orange colors and other muon subdetectors are highlighted in different colors ~\cite{h}.
\label{Fig:Muon system}}
\end{figure*}
\\
This paper presents the first comprehensive report on the efficiency of GE1/1 chambers, including the dependence of efficiency on position within a chamber, pile-up, and the effect of high voltage discharges~\cite{k}. The efficiencies are measured with muons from proton–proton  collisions in Run-3 corresponding to an integrated luminosity of 17.8 fb$^{-1}$ collected during 2023. These results represent the first full-system evaluation since the Run-2 slice test in 2018~\cite{l}, and they validate the readiness of GE1/1 for HL-LHC operation. 

\subsection{Muon System and its Upgrade}
In the configuration of the muon system that was used in Runs 1, 2, and 3 and will continue to be used in future runs, the barrel contains DTs and RPCs, and the endcaps contain RPCs and CSCs. These choices are made based on the anticipated background rates within each region. The muon system ~\cite{m} consists of four stations in the barrel region and four disks in each of the endcaps, effectively covering the region $|\eta|<2.4$. 
\\
\\
The arrangement and positioning of the detectors are determined based on the requirements to effectively cover a wide detection area and to experience varying degrees of radiation exposure. In the barrel region, where the residual magnetic field and the muon flux are both rather low, the DTs~\cite{n} are used. They cover the region $|\eta|<1.2$, delivering precision measurements, and Level-1 (L1) triggering information. In the two endcap regions of the CMS detector, where the magnetic field is strong and non-uniform, and high muon flux prevails, CSCs~\cite{o} are deployed and cover the region $|\eta|<2.4$ providing similar functions. The CSCs are gaseous trapezoidal Multi-wire Proportional Chambers and are characterized by a short drift length, which leads to a fast signal collection. In each endcap, there exist four stations of CSCs strategically oriented perpendicular to the beamline and interspersed among the flux return plates.
\\
\\
In both the barrel and endcap regions, RPCs~\cite{p} are employed to increase redundancy in trigger and position measurement. The RPCs provide fast response with reliable time resolution of $\sim$1.5 ns~\cite{p}, although their position resolution is slightly less precise than the DTs and CSCs. For the Phase 2 upgrade of the CMS muon detector system, three stations, named GE1/1, GE2/1, and ME0 based on GEM technology are to be included in the endcaps of the muon system~\cite{i,j}. 

\section{The CMS GE1/1 GEM Chambers}
\begin{figure*}[htb!]
\centering
\includegraphics[width=7.0cm]{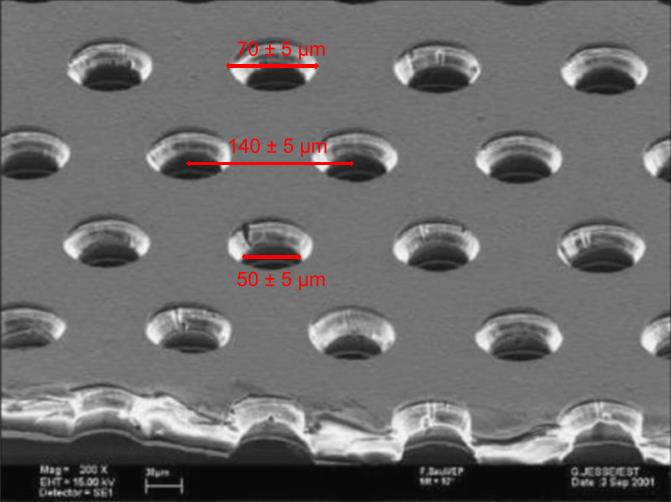}
\includegraphics[width=5.23cm]{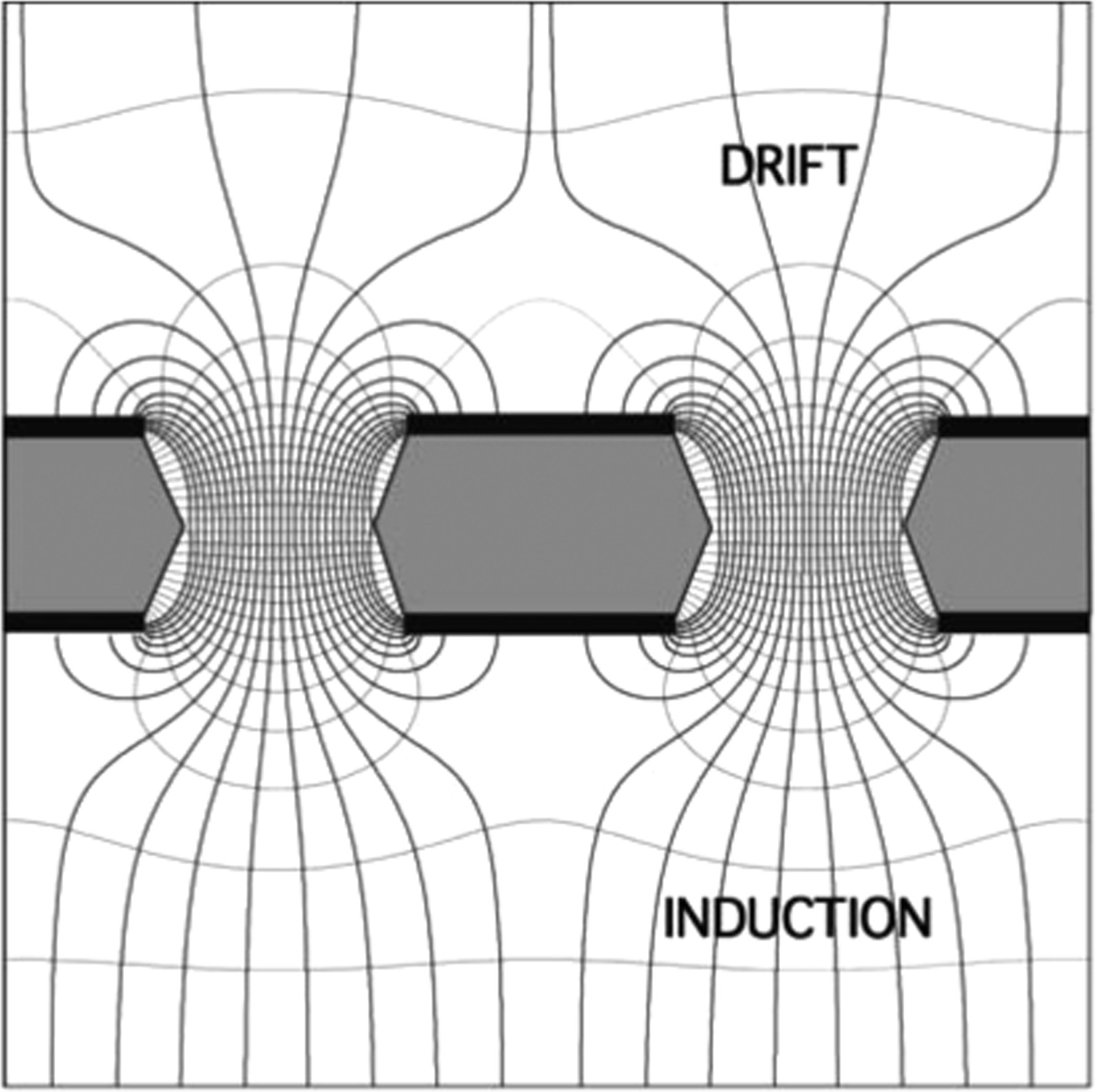}
\caption{Electron microscope image of a typical GEM foil, 50 $\mu$m thick (left). The dimensions in red are specific to the GE1/1 foils. Electric field lines are shown in the region of the holes (right)~\cite{q}.
\label{Fig:GEM foil}}
\end{figure*}

\begin{figure*}[htb!]
\centering
\includegraphics[width=10.5cm]{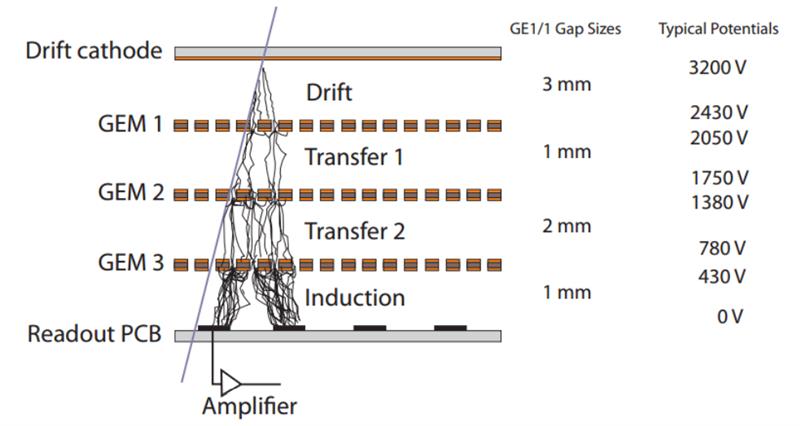}
\caption{Principle of operation of the triple-GEM chamber showing the layout of the drift, transfer, and induction regions.  Typical electric potentials for a GE1/1 chamber are shown adjacent to each foil and PCB surface.}
\label{Fig:triple_GEM_detector}
\end{figure*}

GEM detectors, categorized as Micro Pattern Gas Detectors (MPGDs), utilize microscopic structures to initiate avalanches, effectively multiplying the number of electrons in the process. GEM detectors utilize thin polyimide (Kapton) foils, coated on both sides with copper. As shown in figure \ref{Fig:GEM foil}, the foils are perforated by holes with a surface density of 50-100 per mm$^2$~\cite{q}.
\\
\\
Using photolithography techniques, holes with a double-funnel shape are chemically etched with the holes in a regular hexagonal arrangement. The CMS GEM chambers are constructed using two printed circuit boards containing the gas volume of Ar and CO$_2$ in a ratio 70:30, and in between the boards there is a stack of three GEM foils separated by a few millimeters, immersed in a gas mixture. The gas mixture is ideal for a harsh radiation environment because it lacks polymerizing agents, it is non flammable, and it makes no contribution to the greenhouse effect. A schematic view of the operating principle is shown in figure \ref{Fig:triple_GEM_detector}, which also defines the three types of regions within the triple-GEM chamber.
\\
\\
The drift region lies in the 3 mm gap between the PCB cathode and the first GEM foil. The first transfer region lies between the first and second GEM foils, the second transfer region lies between the second and third GEM foils, and the induction region lies between the third GEM foil and the readout PCB. When a charged particle passes through the gas, free electrons are produced by ionization and are accelerated by a static electric field to move towards a GEM foil. As shown in figure \ref{Fig:triple_GEM_detector} (right) there is a large electric field in the region of a hole in the foil allowing electrons  to acquire enough energy to subsequently ionize gas molecules. This process leads to a cloud of electrons many more in number than the primary number of ionization electrons from passage of the charged particle. The movement of the electron cloud towards the anode strips, induces a signal suitable for electronic readout.

\subsection{Motivation for the GE1/1 Detector Station}
The primary goal of the GE1/1 station is to aid in muon reconstruction and manage level-1 muon trigger rates during HL-LHC operation. High background particle rate makes signal identification difficult and significantly affects detector performance. The main source of background affecting hit rate and occupancy levels in the muon detectors is comprised of neutrons and secondary particles from neutron interactions with matter. The backgrounds complicate the association of individual hits with tracks, posing challenges for event reconstruction for both the muon trigger and the offline analysis.
\\
\\
The addition of the GE1/1 station makes it possible to measure $p_T$ in the muon-only level-1 trigger. Although the magnetic field lines in the endcaps have significant curvature, the bending in the first muon station between GE1/1 and ME1/1 is sufficient to measure $p_T$. This makes it possible to maintain the low-$p_T$ threshold used for LHC operation while maintaining acceptable trigger rates for the HL-LHC. The combination of the GE1/1 and ME1/1 stations also enhances the muon reconstruction for the High-Level Trigger and the offline reconstruction.

\subsection{GEM Chamber Configuration and Operation}
The layout of a GEM chamber, shown in figure \ref{Fig:triple_GEM_detector}, was chosen to optimize its performance. The 3 mm gap between the cathode and first amplification stage provides enough primary ionization charge to achieve efficient particle detection. The smaller gaps, labeled Transfer 1, Transfer 2, and Induction, help to minimize chamber thickness for mechanical advantage and obtain good spatial and time resolutions by minimizing the transverse size of the charge cloud and the drift time through the chamber.
\\
\\    
The GEM foils~\cite{q} are made from a flexible 50 $\mu$m thick polyimide substrate with a 5 $\mu$m thick copper coating on both sides. Holes are etched in the foil in an hexagonal pattern with 140 $\mu$m spacing between holes. As shown in figure \ref{Fig:GEM foil} (right) the holes have a biconical shape with 70 $\mu$m diameter at the surfaces and 50 $\mu$m at the center.
\\
\\
The electric potential differences between the foil surfaces, shown in figure \ref{Fig:triple_GEM_detector}, are chosen so that the electric field strength across each foil, and hence the amplification factor of each foil, is successively smaller in the sequence of foils from 1, to 2, to 3. This configuration minimizes gas discharge formation~\cite{r}. 
The time resolution of a GEM detector is measured to be approximately 5 ns with a signal width of approximately 43 ns~\cite{h}. 

\subsection{GE1/1 System Design}

\begin{figure*}[htb!]
    \centering
    \includegraphics[width=10.5cm]{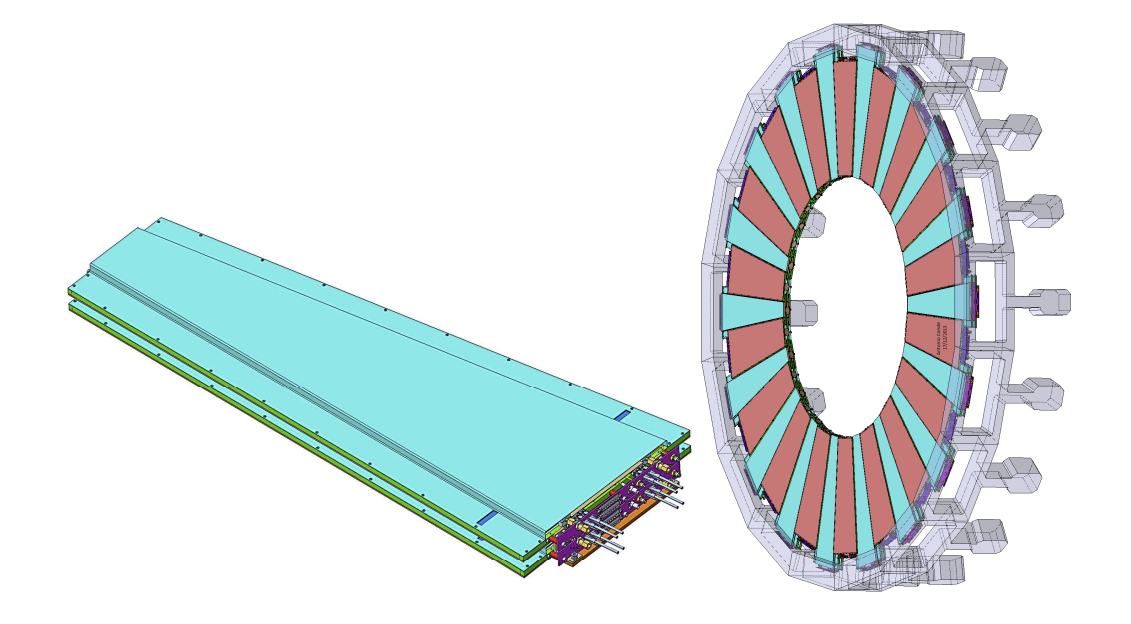}
    \caption{A pair of GE1/1 chambers form a super-chamber (left). Long and short chambers are combined to accommodate the mechanical constraints of the supporting structure~\cite{s}.}
    \label{fig:GEM_chambers}
\end{figure*}

The GE1/1 detection system is designed to provide maximum geometrical acceptance within the CMS envelope, high rate capability of 10 kHz/cm$^2$ or better, 97$\%$ efficiency for detecting minimum ionizing particles, angular resolution of 300 $\mu$rad or better, and uniform gain of 15$\%$ or higher across individual chambers and between them~\cite{h,s}. These properties remain stable even after accumulating 200 mC/cm$^2$ of integrated charge.
\\
\\
The GE1/1 system consists of 144 trapezoidal-shaped chambers organized into 72 super-chambers (SC). Each SC consists of two chambers mounted together, as shown in figure \ref{fig:GEM_chambers} (left). The two chambers are held together by an aluminum plate and L-brackets, with Layer-1 facing the interaction point and Layer-2 facing in the opposite direction. This setup enables increased efficiency and angular precision. The two chambers that are chosen for a given SC have similar gas gain at a fixed voltage. 
\\
\\
On each endcap, 36 SCs form a complete azimuthal ring, as shown in figure \ref{fig:GEM_chambers} (right). To accommodate the mechanical constraints of the supporting structure, there are two types of chambers with different radial lengths. The longer and shorter types, GE1/1-L and GE1/1-S, cover regions 1.55 < |$\eta$| < 2.18 and 1.61 < $|\eta|$ < 2.18, respectively~\cite{t}. There are 18 long and 18 short SCs in each ring. The SCs have been installed in the slots originally intended for RPC chambers, within the gap located between the hadron calorimeter and the CSC ME1/1 chambers situated in the YE1 "nose". 
\\
\\
Each GEM chamber operates as an independent unit within the readout electronics system. For strip readout, the area of a chamber is divided into 24 sectors with 3 divisions in $\phi$ and 8 divisions in $\eta$. Each sector contains 128 strips read out by a single 128 channel VFAT3~\cite{u} front-end chip. Data collected from VFAT3 chips are routed via the GEM Electronics Board to the opto-hybrid field-programmable gate array board, located near the outer radius of the chamber. The 8 VFAT3 chips, corresponding to the 8 sectors covering the same region in $\phi$, are organized into "columns". The OptoHybrid board drives 3 optical fibers, each transmitting tracking data for one column to the backend electronics located off-chamber in the experimental cavern. The backend electronics communicate with the CMS Data Acquisition system located in the service cavern.

\section{GE1/1 Detection Efficiency}
This section describes the method used to evaluate the efficiency of the GE1/1 detectors. Efficiency is determined for each individual chamber of each super-chamber as well as for the chamber region readout by individual VFAT chips. This fine-grained analysis provides insight into the dependence of efficiency on readout chip performance, operating voltage, and mechanical effects such as foil deformation.
Also measured is the dependence of efficiency on pile-up, the number of interaction vertices per beam bunch crossing. A sample of high-purity muons from Z boson decay was used to measure GE1/1 efficiency.
In the sub-sections below, the methods are described used to find muon tracks and determine efficiency, followed by the results on efficiency.

\begin{figure}[htb!]
    \centering
    \includegraphics[width=4.5cm]{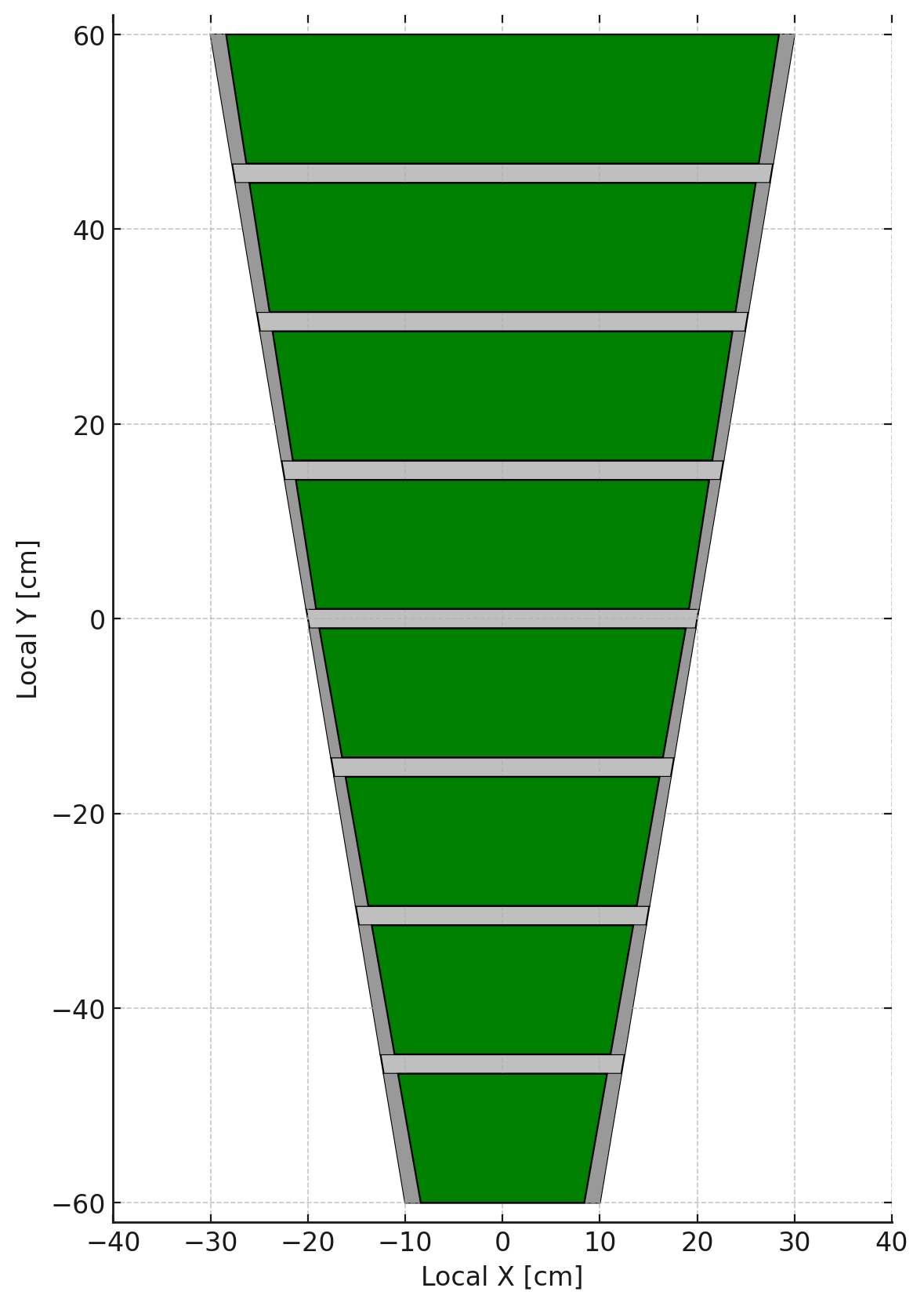}
    \caption{GE1/1 detector $\eta$-partitions and acceptance region.}
    \label{fig:active_reg}
\end{figure}

\subsection{Fiducial Geometry}
Each chamber is segmented into eight $\eta$ partitions, as shown in figure \ref{fig:active_reg}. Fiducial regions are defined to lie at least 0.5 cm from a partition boundary and at least 2 mrad away from a chamber edge. This insures that inactive areas of a chamber are excluded from the efficiency calculation.

\subsection{Algorithm for Muon Reconstruction}
The pool of candidate muon tracks used for this study are reconstructed starting with the standard algorithm for physics data taking in Run 3, which utilizes hits from GE1/1, the other detectors in the muon endcap (CSC and RPC), and the tracker. To measure GE1/1 efficiency, the tracks are subsequently re-reconstructed without use of GE1/1. 
Standard muon track reconstruction~\cite{v} involves three types of tracks: tracker, standalone (STA), and global. 
\\
\\
Tracker muons are reconstructed using an iterative approach, where tracks are extrapolated to the muon system. A tracker track is identified as a tracker muon if it matches to at least one muon segment in the muon system. This matching process is done in the transverse plane, taking into account the uncertainty in mechanical alignment.
\\
\\
STA muon tracks are reconstructed using information exclusively from the muon subdetectors (CSC, GEM and RPC) through a Kalman filter technique. The process begins with seed segments from the CSC subdetectors, which are then used to collect information along the muon’s trajectory.
\\
\\
Global muon tracks are created by combining STA muon tracks with tracker muon tracks. The process begins by matching their track parameters on a common reference surface, followed by a Kalman filter fit that incorporates information from both the tracker and the STA muon tracks.  The criteria for global muon tracks are given in Table 1. 

\begin{table}[h!!!]
\begin{center}
\caption{Muon Identification Criteria}
\vspace{3mm}
\begin{tabular}{|l|c|c|}
\hline
Item & Requirement\\\hline\hline
Muon Reconstruction & Reconstructed as a global muon with particle flow (PF) algorithm\\
\hline
Tracker Track & Hits from at least 6 layers of the tracker,\\
              & including 1 pixel hit,\\
\hline
Muon Segment Matching & Segment matching in at least \\
                              & 2 muon stations\\
\hline
Global Muon Fit & $\chi^2/\text{dof}<$10 and at least 1 hit from \\
                                & the muon system\\
\hline
Vertex Compatibility & Transverse impact parameter: $|d_{xy}| < 0.2$ cm\\
                      & Longitudinal impact parameter: $|d_z| < 0.5$ cm\\
\hline
\end{tabular}
\label{tab:Muon Id}
\end{center}
\end{table}

\noindent
We use tight muon identification criteria which ensures selection of high purity, well reconstructed muons. Tight muon identification is crucial for rejecting muons originating from decays-in-flight and hadronic punch-through. Each muon must be reconstructed as a global muon with a particle flow (PF) algorithm. The associated tracker track is required to have hits in at least six layers, including at least one pixel hit. Moreover, the muon must be reconstructed as both a tracker muon and a global muon, with consistent segment matching in at least two muon stations, ensuring accuracy and reliability in data analysis. The global muon fit must have $\chi^{2}/\text{dof} \textless$ 10, and must include at least one hit from the muon system. Additionally, a tight muon must be compatible with the primary vertex, having a transverse impact parameter ${|d_{xy}|\textless}$ 0.2 cm and a longitudinal impact parameter 
${|d_z|\textless}$ 0.5 cm. 

\subsection{Back-Propagation Technique}
\begin{figure}[htb!]
    \centering
    \includegraphics[width=12.0 cm]{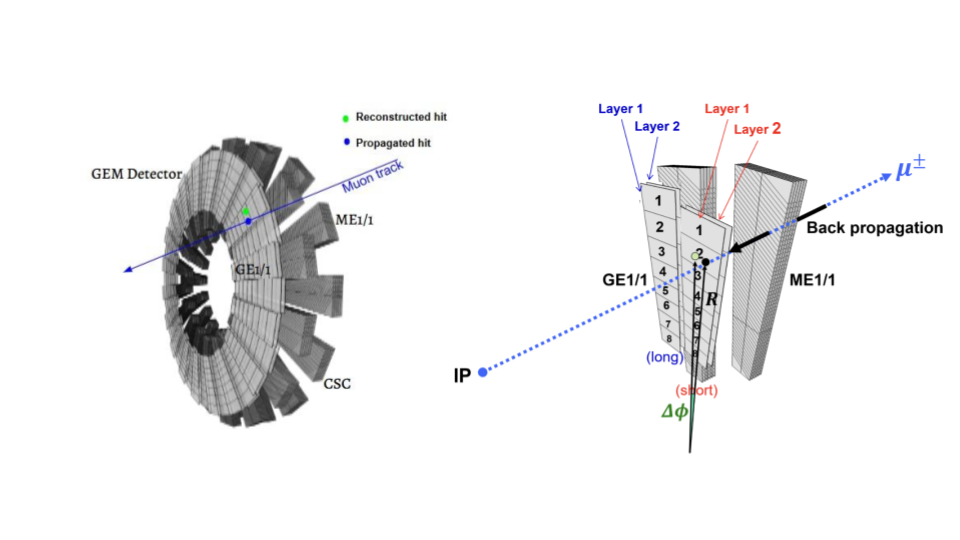}
    \caption{Back propagation procedure to GE1/1 from ME1/1 showing entire ring of chambers (left) and detailed view of two chambers in each ring (right). For both views, the projected hit is shown in blue and the GE1/1 hit is shown in green. The matching procedure is based on the arc length ($R\Delta\phi$) between the GE1/1 and projected hits, shown in the detailed view (right) ~\cite{x}.}
    \label{fig:propHit_demo}%
\end{figure}

\begin{figure}[htb!]%
    \centering
    \includegraphics[width=7.0cm]{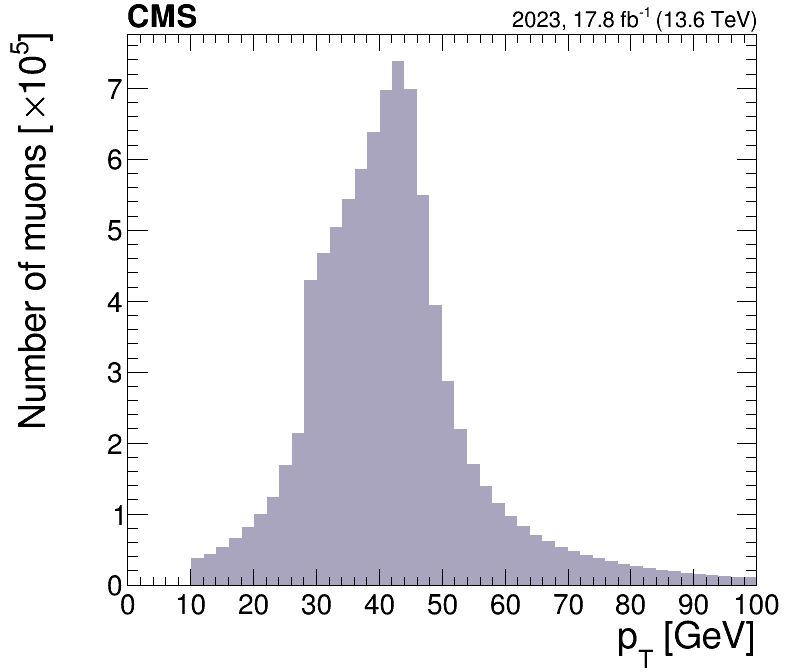}
    \caption{The ${p_T}$ distribution of all global muon tracks used in the efficiency analysis. One of the selection criteria is ${p_T}$ > 10 GeV.}%
    \label{fig:pt_dist}%
\end{figure}

To measure GE1/1 efficiency, a global muon track is re-fit with the GE1/1 hits removed from the STA track. A back-propagation~\cite{w} algorithm is used to project global muon tracks (without GEM hits) back through the muon system to predict their intersection with GE1/1. 
Muons are selected from events collected with a trigger requiring an isolated muon with $p_T>24$ GeV, providing a well-measured sample of reconstructed muons. Events are selected only from periods with normal GE1/1 operating conditions.
\\
\\
The reconstructed track (without GE1/1 hits) is propagated backwards from the outermost muon station, through the muon system, towards the GE1/1 detectors, providing an estimate of the expected hit positions. If a reconstructed GE1/1 hit falls within $\pm$4 cm of the expected position in $R\Delta\phi$, it is considered a match. Global muon tracks are required to point inside the GE1/1 acceptance, have $p_T>10$ GeV, and contain at least one CSC ME1/1 segment. The back propagation procedure is depicted in figure \ref{fig:propHit_demo}, and the ${p_T}$ distribution of global muon tracks used in the efficiency analysis is shown in figure \ref{fig:pt_dist}.

\vspace{-0.5cm}
\subsection{Efficiency Calculation and Results for Each Chamber}
\begin{figure}[htb!]
    \subfloat{{\includegraphics[width=6.5cm] {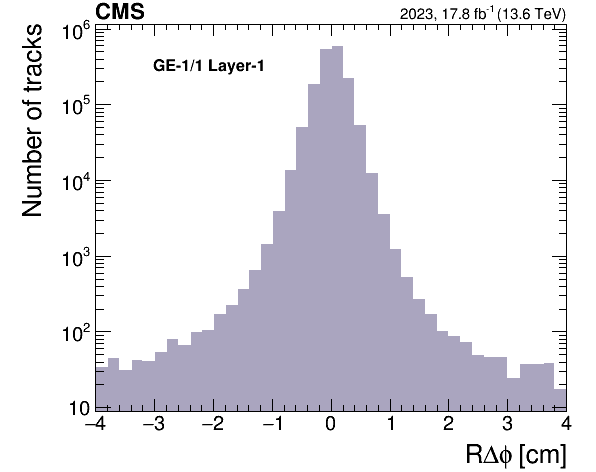}}}%
    \qquad
    \subfloat{{\includegraphics[width=6.5cm]{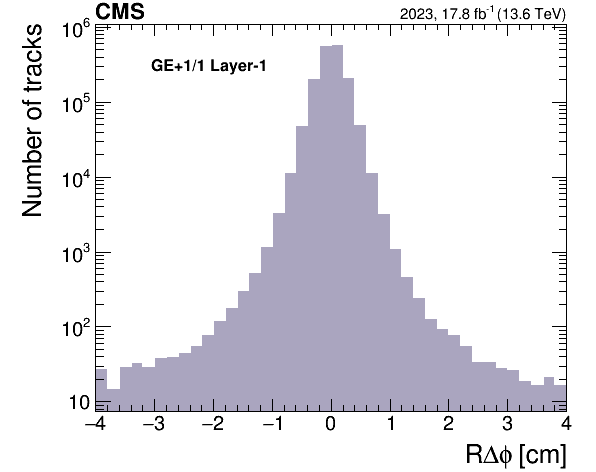}}}
    \newline
    \subfloat{{\includegraphics[width=6.5cm] {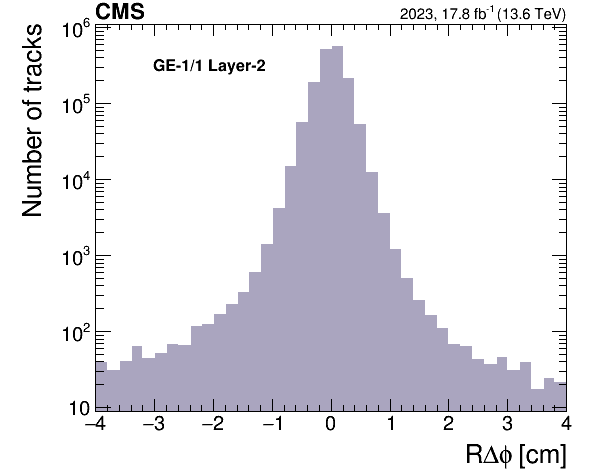}}}%
    \qquad
    \subfloat{{\includegraphics[width=6.5cm]{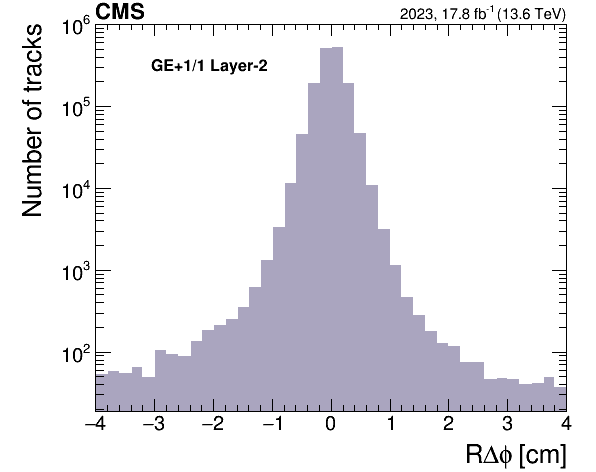}}}
\caption{Residual $R\Delta\phi$ distribution used to match GE1/1 hits with back-propagated muon tracks. Distributions are shown for entire rings of chambers: negative endcap (left column), positive endcap (right column), layer 1 (top row), layer 2 (bottom row). Here and in subsequent plots, GE-1/1 and GE+1/1 indicate, respectively, the negative side and positive  side endcaps.}
\label{fig:residual_dist}
\end{figure}

\begin{figure}[htb!]
\includegraphics[width=14cm]{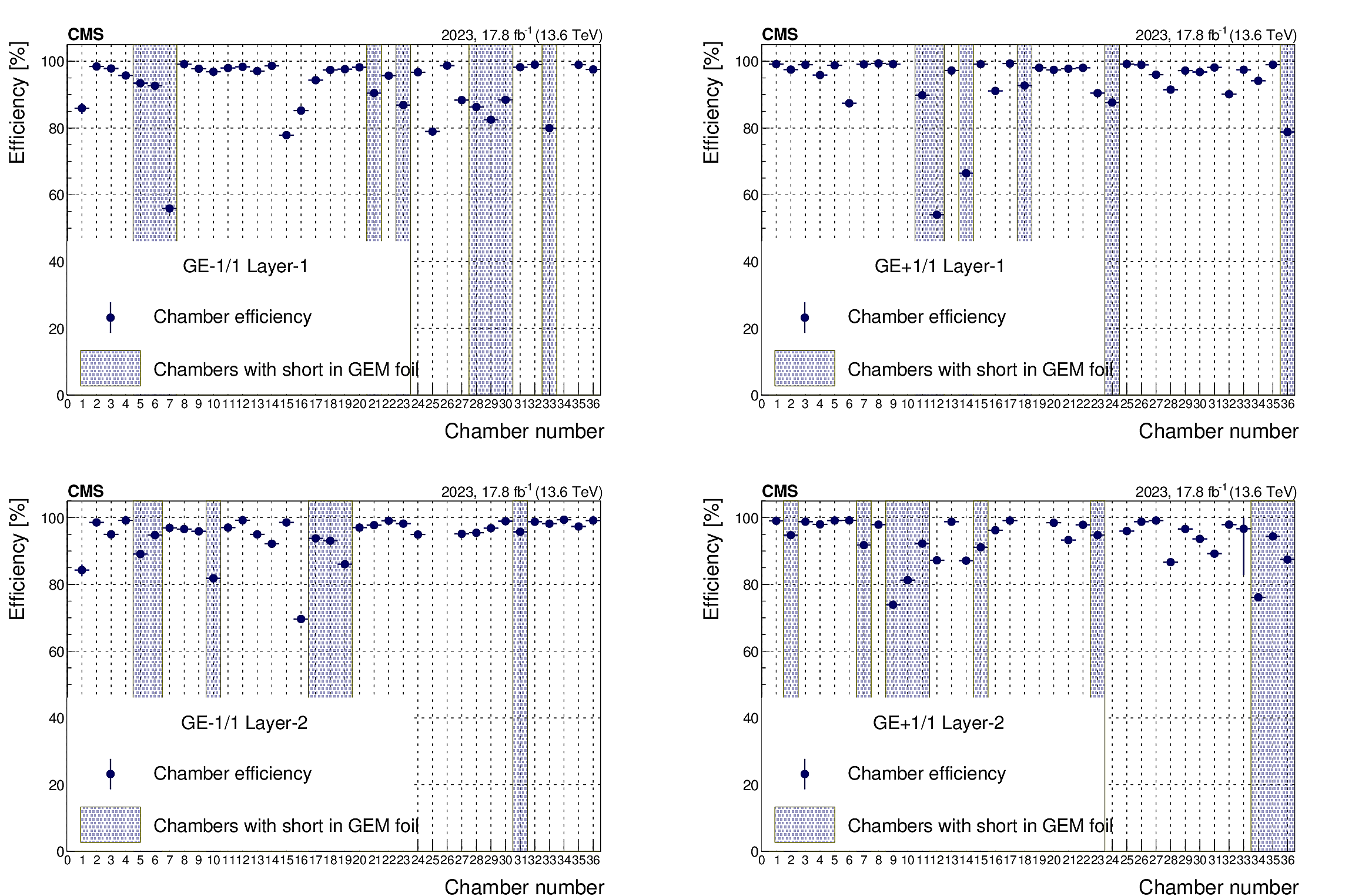}%
\centering
\caption{Efficiency is shown with a dot for each of 137 out of 144 chambers. There were 7 chambers that were turned off during data taking. The left (right) figure shows the efficiency of negative (positive) endcaps while the top (bottom) plots show the efficiency of Layer-1 (Layer-2).  There are 32 chambers with at least one short; these are highlighted by shading.}
\label{fig:eff_per_cham}
\end{figure}

\begin{figure}[htb!]
\includegraphics[width=0.65\linewidth]{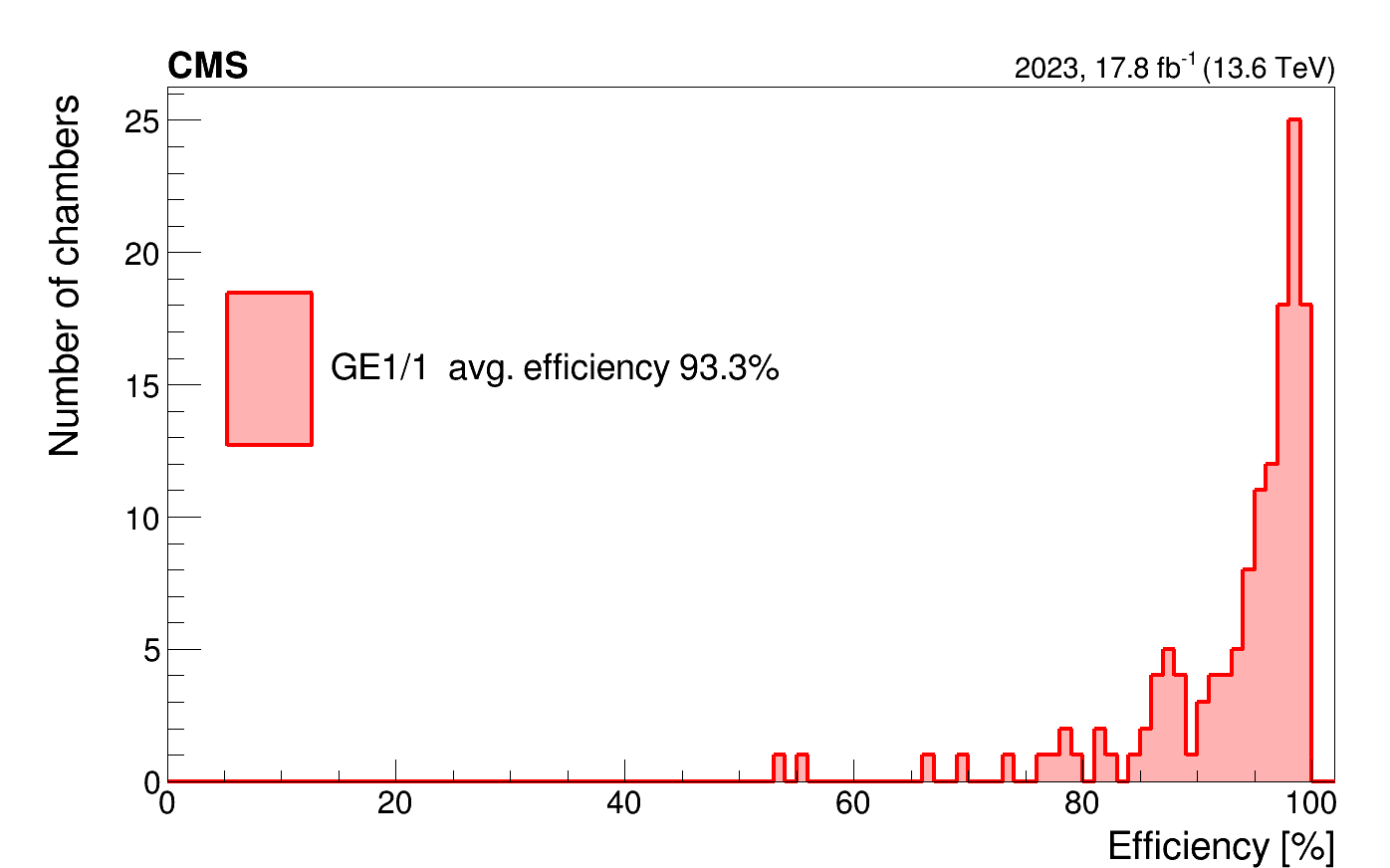}
\centering
\caption{Efficiency of 137 chambers. An average efficiency of 93.3$\%$ is observed for all the chambers.}
\label{fig:eff_dist}
\end{figure}

Efficiency is defined as the ratio of the number of tracks with matched hits to the total number of tracks. Examples of the distribution of $R\Delta\phi$ residuals used in the matching procedure are shown in figure \ref{fig:residual_dist} for four individual chambers. All four distributions have prominent peaks, well-centered in the required $\pm4$ cm range.
\\
\\
The efficiencies for each of 137 chambers are shown in figure \ref{fig:eff_per_cham}. There were 7 chambers that were turned off during data taking. The majority of the chambers show efficiency above 95$\%$, though there are exceptions among a subset of chambers identified by the presence of at least one or more electrical shorts or those operating at non-optimal working points. A histogram of the 137 efficiency values is shown in figure \ref{fig:eff_dist}.
The average efficiency is 93.3$\%$. Only a small fraction of chambers show efficiency less than 90$\%$.

\subsection{VFAT-Level Efficiency}
The efficiencies of the chambers for the regions covered by all individual VFATs are shown in figure 11 using a color-coded scale. It is evident that most VFATs are associated with high-efficiency (green), while there are regions that are excluded (black) because of operational issues. The most common problem~\cite{y} was due to outgassing of adhesive materials on opto-hybrid boards causing failure of the optical connection. A mitigation procedure providing enhanced cooling is planned for implementation during the upcoming long shutdown. Regions with reduced efficiency due to short circuits on the GEM foils and operation at lower than nominal HV are discussed in later sections.


\begin{figure}[htb!]
    \subfloat{{\includegraphics[width=6.5cm]{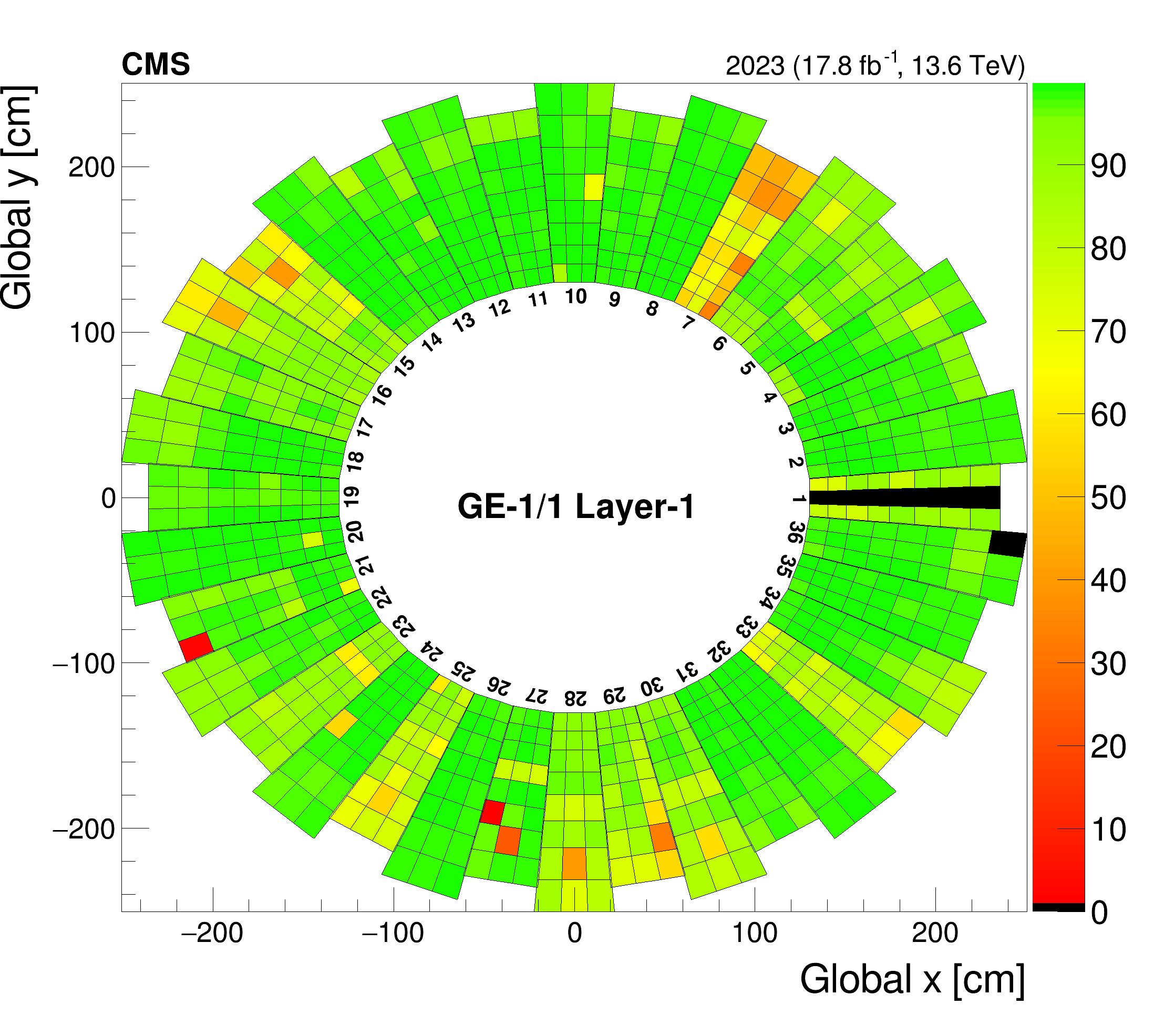}}}
    \qquad
    \subfloat{{\includegraphics[width=6.5cm]{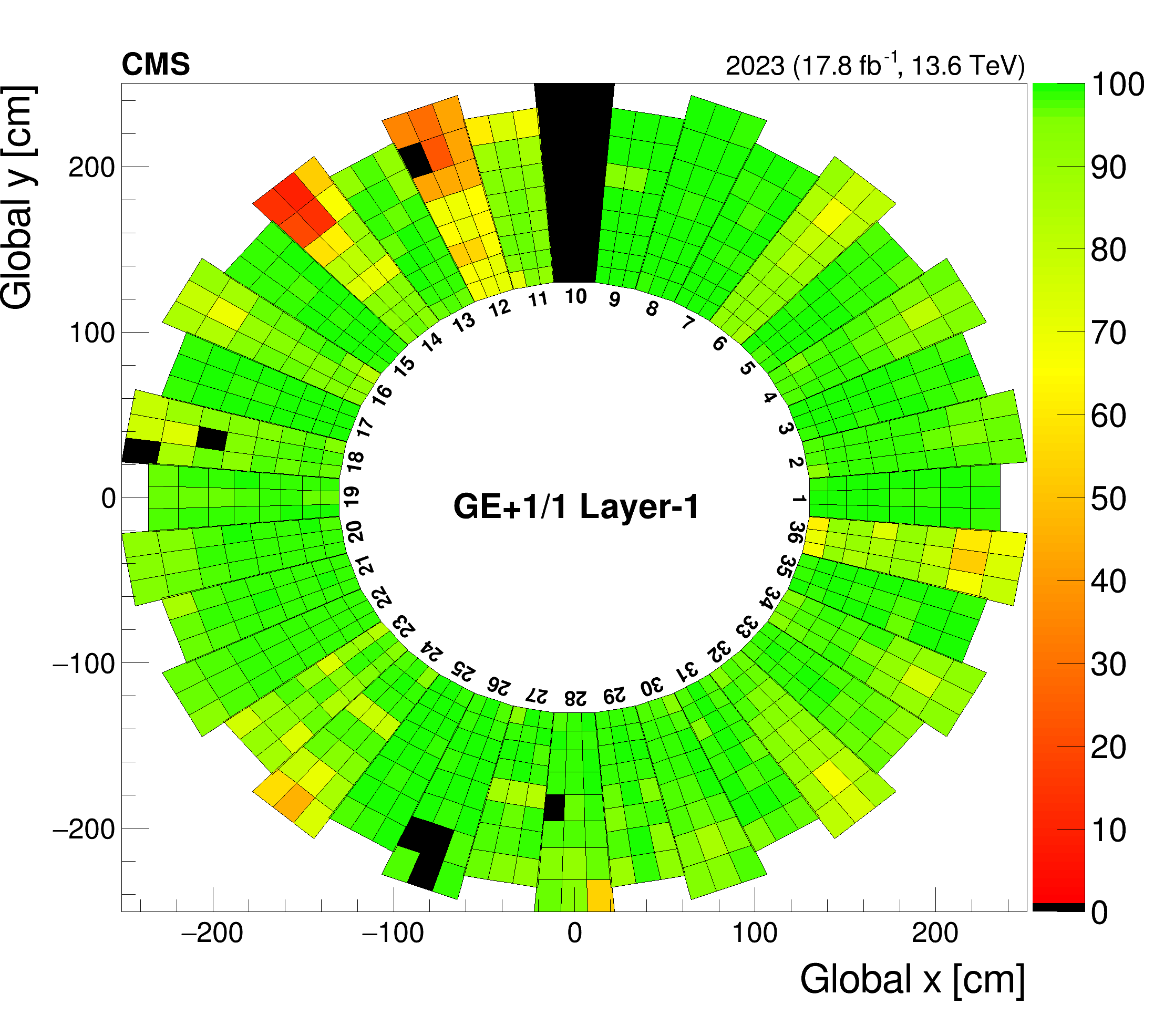}}}
    \newline
    \subfloat{{\includegraphics[width=6.5cm]{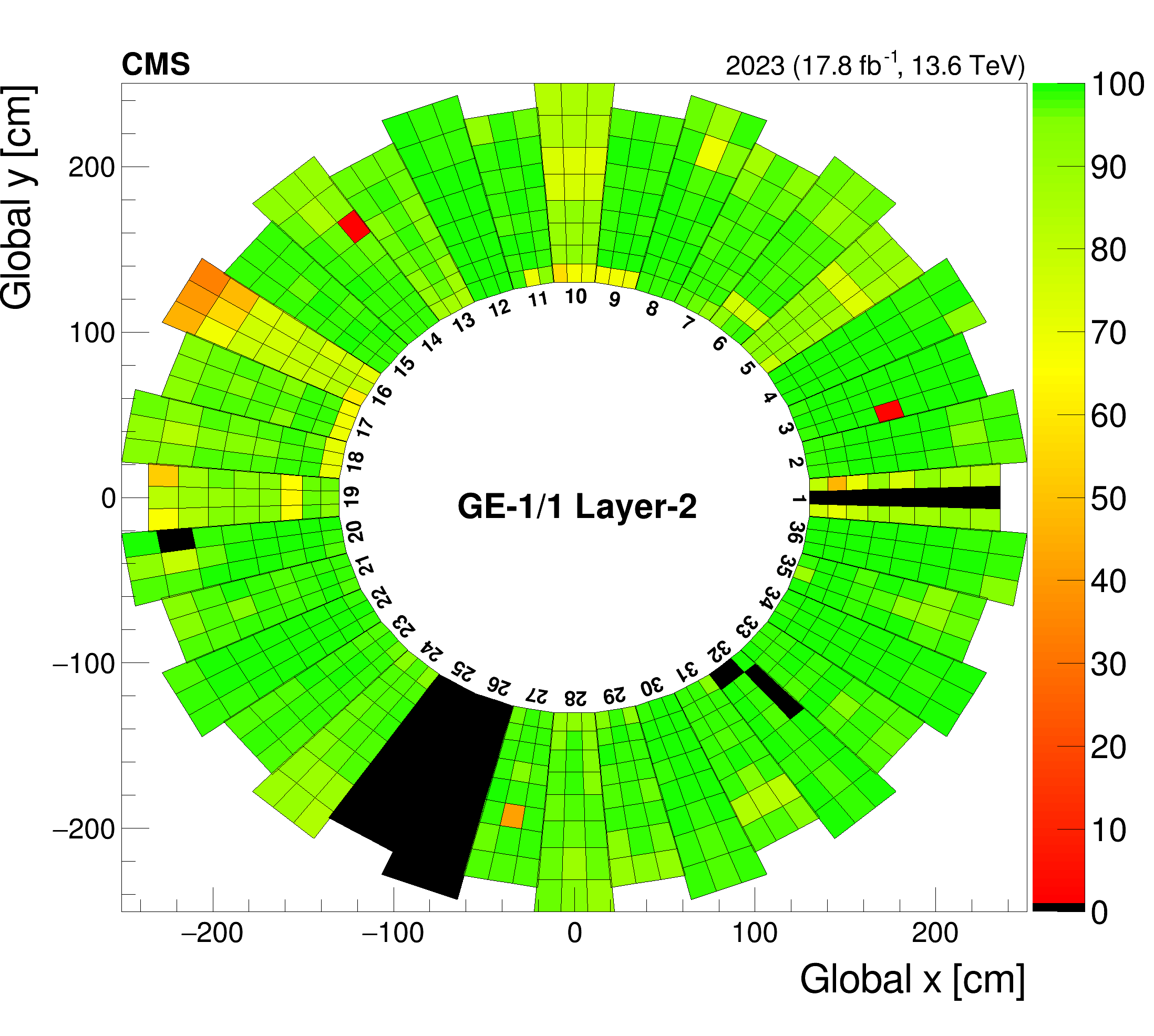}}}
    \qquad
    \subfloat{{\includegraphics[width=6.5cm]{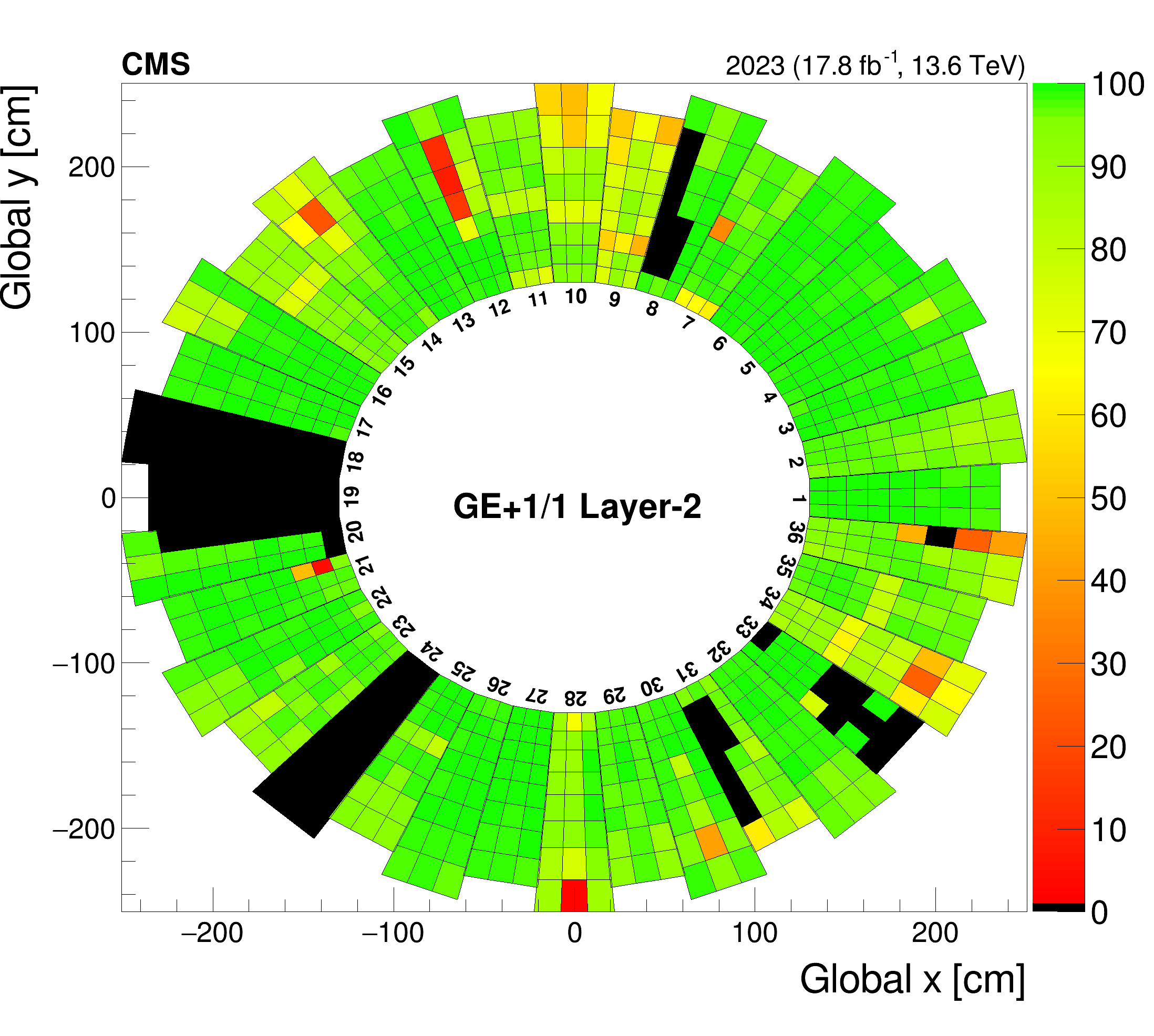}}}

\caption{Efficiencies for the regions covered by all individual VFATs. Regions are excluded (black) because of operational problems described in the text.}
\label{fig:effpervfat}
\end{figure}

\subsection{Pile-Up Dependence}
\begin{figure}[htb!]
\includegraphics[width=0.65\linewidth]{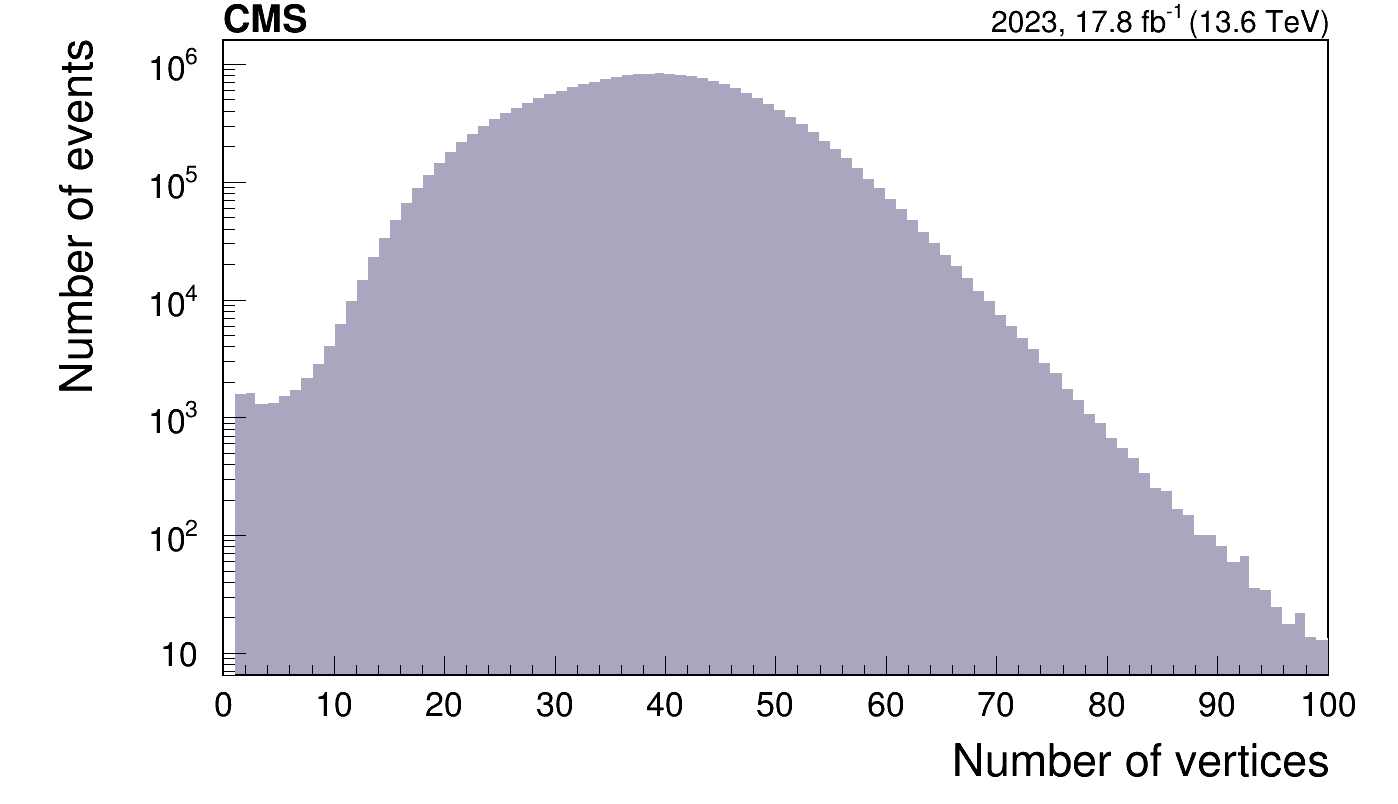}
\centering
\caption{Distribution of number of reconstructed vertices per event.}
\label{fig:vrtx_dist}
\end{figure}

\begin{figure}[htb!]
\includegraphics[width=0.50\linewidth]{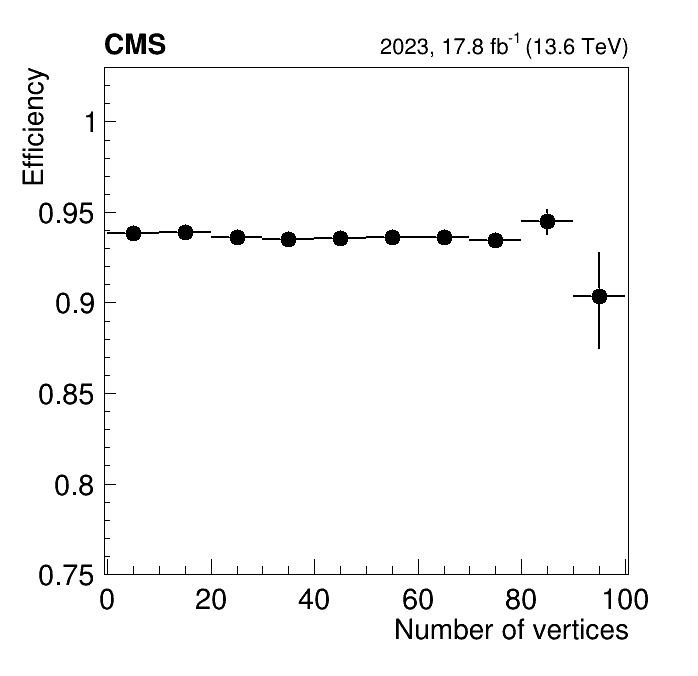}
\centering
\caption{Efficiency versus number of vertices per event.  Chambers in both layers of both endcaps are included.}
\label{fig:effnvtx}
\end{figure}

The term “pile-up” refers to the number of collisions per bunch crossing. A measure of pile-up is the number of reconstructed vertices per event, whose distribution is shown in figure \ref{fig:vrtx_dist} for the full data sample.  As shown in figure \ref{fig:effnvtx}, efficiency is found to be independent of the number of reconstructed vertices per event, hence insensitive to pile-up.

\subsection{High Voltage Instabilities and Short Circuits} 
HV instabilities can result from sudden electrical discharges, triggered by high particle flux, elevated gas gain, or local imperfections such as copper-edge defects and dust. These discharges can produce transient or permanent shorts between electrodes, producing high currents that may thermally damage the GEM foil. Since such shorts can appear or disappear during operation, affected chambers are often operated at reduced high voltage to prevent further deterioration. Out of 108 chambers in both endcaps operated at nominal HV, 30 had shorts. Their impact on detector performance is significant because the efficiency of chambers without shorts is $\sim$96$\%$, whereas those with one short have efficiency of $\sim$89$\%$. Chambers with two shorts have efficiency of $\sim$79$\%$, because multiple defective sectors reduce active area as well as limit the achievable gain. Additional shorts do not always lead to a proportional decrease in efficiency, since multiple shorts often occur in nearby regions. For example, chambers with three shorts have an average efficiency of $\sim$69$\%$, while some chambers with four shorts have efficiency as low as $\sim$54$\%$. These results demonstrate that shorts lead to significant reductions in efficiency and should be avoided by operation with conservative values for HV. 

\subsection{Efficiencies of Chambers Without Shorts Operated at Nominal HV}

\begin{figure}[htb!]
\includegraphics[width=0.60\linewidth]{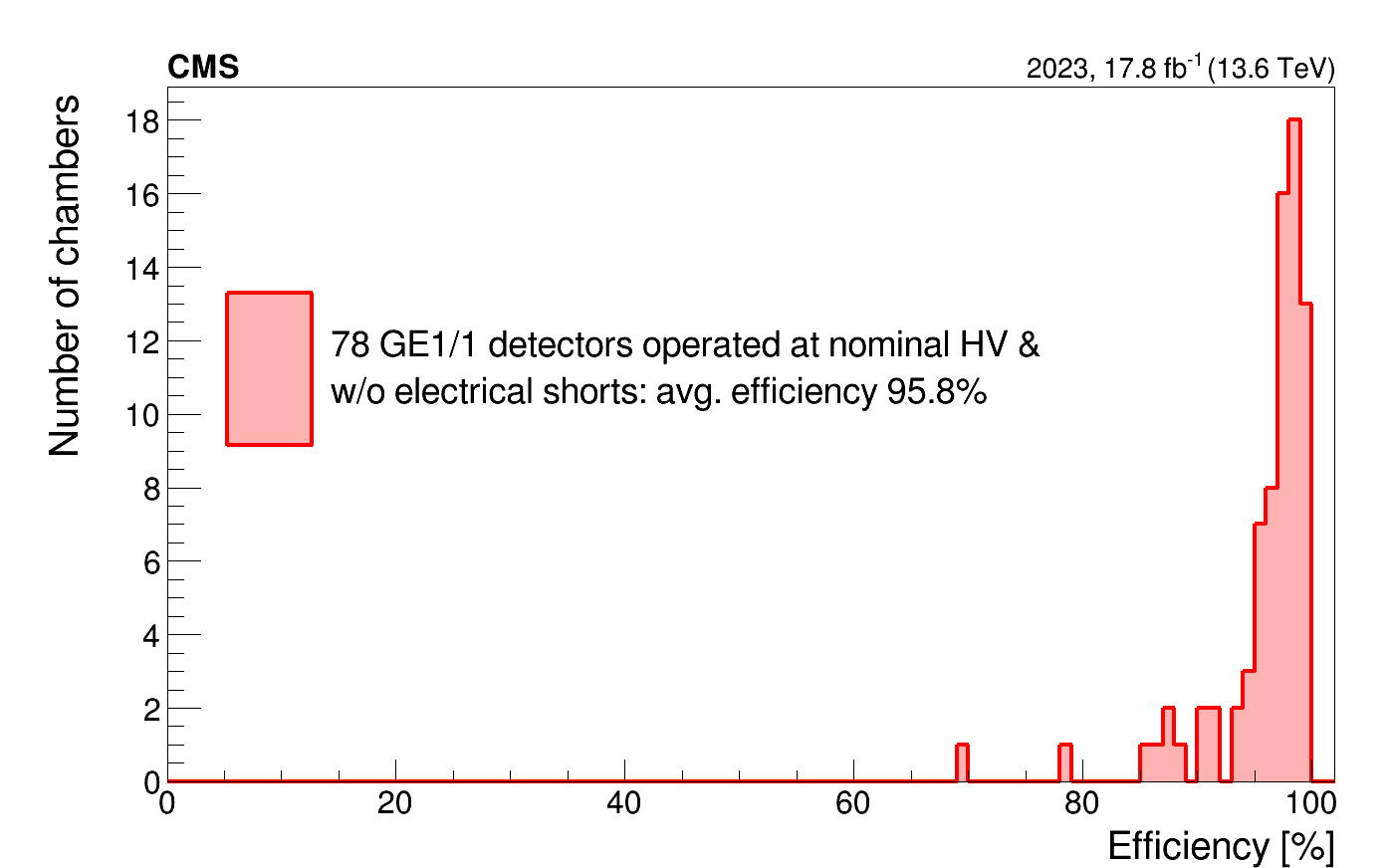}
\centering
\caption{Efficiencies of 78 optimal chambers.}
\label{fig:eff_dist_good_chambers}
\end{figure}

\begin{figure}[htb!]
\includegraphics[width=0.50\linewidth]{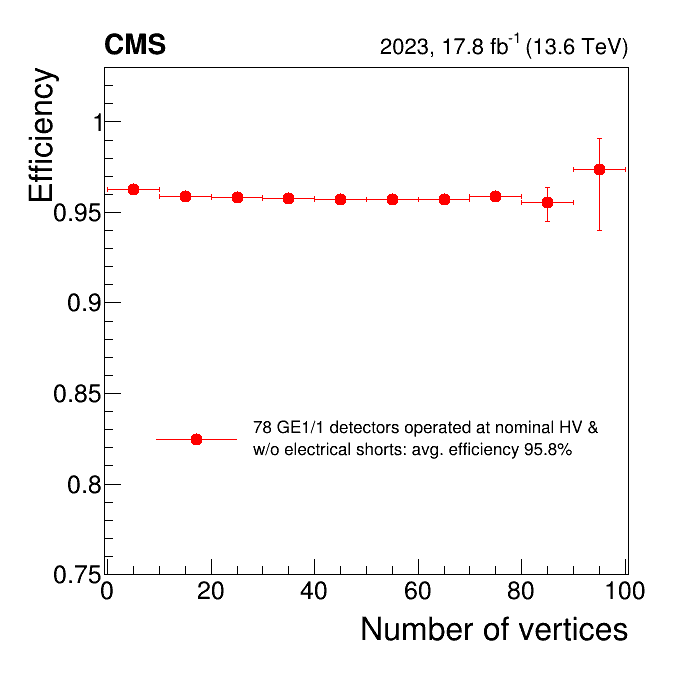}
\centering
\caption{Efficiency vs. the number of vertices for good chambers across all the layers and both endcaps.}
\label{fig:effnvtx_goodchambers}
\end{figure}

To assess the potential efficiency performance of the GE1/1 system using only “optimal” chambers, a set of chambers were selected that were operated at nominal HV, had no electrical shorts, and no evidence of mechanical defects such as printed circuit board warping.
There are 78 out of 137 chambers that meet these requirements, located in the endcaps and layers as detailed in table ~\ref{tab:good_chambers}. The distribution of chamber efficiencies is shown in figure~\ref{fig:eff_dist_good_chambers}. The average efficiency is $\sim$96$\%$, satisfying the performance benchmark in the GEM Technical Design Report ~\cite{i}. For the set of optimal chambers, the efficiency is independent of the number of reconstructed vertices, as shown in figure~\ref{fig:effnvtx_goodchambers}. 

\begin{table}[htbp]
\centering
\caption{Selection information for optimal chambers}
\vspace{3mm}
\begin{tabular}{|l|c|c|c|c|c|}
\hline
\begin{tabular}{c}
Endcap \\ and \\ Layer
\end{tabular} &
\begin{tabular}{c}
Total \\ operational \\ chambers
\end{tabular} &
\begin{tabular}{c}
No. of \\ chambers \\ operated at \\ nominal HV (A)
\end{tabular} &
\begin{tabular}{c}
No. of \\ chambers \\ with at least \\ one short (B)
\end{tabular} &
\begin{tabular}{c}
No. of \\ Good \\ chambers at \\ nominal HV \\ (A $-$ B)
\end{tabular} \\
\hline
GE$-$1/1 L1 & 36 & 26 & 9 & 17 \\
\hline
GE$-$1/1 L2 & 34 & 25 & 7 & 18 \\
\hline
GE$+$1/1 L1 & 35 & 31 & 6 & 25 \\
\hline
GE$+$1/1 L2 & 32 & 26 & 8 & 18 \\
\hline
\textbf{Total} & \textbf{137} & \textbf{108} & \textbf{30} & \textbf{78} \\
\hline
\end{tabular}
\label{tab:good_chambers}
\end{table}

\section{Summary and Conclusions}
The first comprehensive study on the efficiency of GE1/1 chambers is presented using 2023 p-p collision data. 
We measure an average efficiency per chamber of $\sim$93.3$\%$. The efficiency of the chambers operated at nominal HV and having no shorts is $\sim$96$\%$. The efficiency is presented for small regions across the surface of the chambers corresponding to individual readout chips, with most regions showing high efficiency.
The chamber efficiencies are found to be independent of pile-up. To mitigate the inefficiencies caused in part by failure of optical fiber connections, all chambers will be extracted and refurbished during Long Shutdown 3 in the years 2026-2030.

\section*{Acknowledgments}
\vspace{-5pt}
We gratefully acknowledge support from CERN, FRS-FNRS (Belgium), FWO-Flanders (Belgium), BSF-MES (Bulgaria), MOST and NSFC (China), BMBF (Germany), DAE (India), DST (India), INFN (Italy), NRF (Korea), and DOE (USA).






\end{document}